\documentclass[amsmath,floatfix,twocolumn,superscriptaddress]{revtex4-1}
\usepackage{subfigure}
\usepackage{siunitx}
\usepackage{amssymb}
\usepackage{amsmath}
\usepackage{graphicx}
\usepackage{array}
\usepackage{dcolumn}
\usepackage{psfrag}
\usepackage{bm}
\usepackage{color}
\usepackage{multirow}

\begin{document}

\title{Stabilization of Metallic, Excitonic Insulator, and Superionic Phases in Helium--Rare Gas Compounds at Sub-Terapascal Pressures}

\author{Cong Liu}
    \affiliation{Physics Department, Universitat Polit\`{e}cnica de Catalunya, Campus Nord, Jordi Girona 1--3, 
    08005 Barcelona, Spain}

\author{Jordi Boronat}
    \affiliation{Physics Department, Universitat Polit\`{e}cnica de Catalunya, Campus Nord, Jordi Girona 1--3, 
    08005 Barcelona, Spain}

\author{Claudio Cazorla}
\email{claudio.cazorla@upc.edu}
    \affiliation{Physics Department, Universitat Polit\`{e}cnica de Catalunya, Campus Nord, Jordi Girona 1--3, 
    08005 Barcelona, Spain}
    \affiliation{Research Center in Multiscale Science and Engineering, Universitat Polit\`{e}cnica de Catalunya, 
    Campus Diagonal-Bes\`{o}s, Av. Eduard Maristany 10--14, 08019 Barcelona, Spain}

\begin{abstract}
\textbf{Abstract.}~Helium and rare gases (RG: Ne, Ar, Kr, Xe) are typically considered chemically inert, yet under 
	the extreme pressures of planetary interiors they may form compounds with unexpected properties. Using crystal 
	structure prediction and first-principles calculations, we mapped the phase diagram of binary He--RG systems up 
	to $1$~TPa. We identify several previously unknown stoichiometric compounds that are both thermodynamically and 
	vibrationally stable at sub-terapascal pressures, within the reach of modern high-pressure experiments. In 
        particular, AHe$_{2}$ and AHe (A: Ar, Kr, Xe) adopt previously unreported orthorhombic, hexagonal and cubic phases 
	that remain stable over wide pressure ranges. We further find that He--Xe systems host metallic and excitonic 
	insulator phases at pressures nearly an order of magnitude lower than those required for pure helium, offering a 
	pathway to realize these exotic quantum states experimentally. Finite-temperature simulations also reveal superionic 
	He--Xe phases, in which helium ions diffuse either anisotropically or isotropically depending on the host lattice. 
	These findings constitute the first prediction of helium-based systems that combine metallicity and superionicity, 
	with profound implications for energy transport and planetary dynamo processes. Overall, our results demonstrate that 
	mixing helium with heavier rare gases provides an effective strategy to stabilize metallic, excitonic insulator, 
	and superionic phases at experimentally accessible pressures, opening new research directions for condensed matter 
	physics and planetary science. 
	\\

{\bf Keywords:} high-pressure physics, first-principles calculations, helium, rare-gases chemistry, exotic condensed-matter phases 

\end{abstract}

\maketitle

\section{Introduction}
\label{sec:intro}
Helium (He) and other rare gases (RG) such as neon (Ne), argon (Ar), krypton (Kr), and xenon (Xe) play a pivotal role in 
Earth and planetary sciences, particularly under extreme pressure conditions characteristic of planetary interiors. Despite 
their chemical inertness at ambient conditions, these elements can exhibit remarkable physical and chemical behavior at 
high pressures, including the formation of unexpected compounds and phase transitions that may influence the structure and 
evolution of giant planets and exoplanets. For example, He's potential to form stable compounds with alkali and alkaline 
earth metals at megabar pressures has profound implications for the understanding of deep interiors of gas giants like Jupiter 
and Saturn, where helium segregation and immiscibility may contribute to thermal evolution and energy transport processes 
\cite{intro1,intro2,intro3}. Similarly, the high-pressure behavior of Xe and its apparent depletion in Earth's atmosphere 
has sparked ongoing debates about its retention and possible incorporation into deep-Earth materials \cite{intro4,intro5}. 
Understanding the reactivity and phase behavior of rare gases under such conditions is therefore essential for developing 
accurate models of planetary formation, composition, and dynamics.

Recent theoretical investigations have revealed that solid helium subjected to extreme compressions of several terapascals 
($1$~TPa = $1,000$~GPa) undergoes a remarkable sequence of electronic phase transitions \cite{cong23}. Specifically, using 
a combination of density functional theory (DFT) and many-body perturbation theory, it was shown that ultra-compressed 
He first stabilizes a bulk excitonic insulator (EI) phase --an exotic state of matter where electron-hole pairs form 
spontaneously without optical excitation-- and then transitions into a superconductor with a critical temperature of 
approximately $70$~K at $100$~TPa \cite{cong23}. These phenomena are of fundamental interest, with potential implications 
for understanding the electronic and thermal evolution of helium-rich white dwarfs and other astrophysical bodies, where such 
extreme pressures may naturally occur. 

Nevertheless, the pressures required to stabilize EI and superconducting phases in helium remain well beyond the reach of current 
static and dynamic compression experiments on Earth, which are typically limited to a few TPa \cite{tpa1,tpa2}. This limitation 
motivates the search for alternative pathways to access similar quantum phases at more moderate and experimentally attainable 
pressures, ideally within the sub-TPa range (hundreds of GPa). A promising strategy is to explore chemically engineered helium 
compounds incorporating heavier RG, as the resulting chemical pressure may significantly modify their structural and electronic 
properties at reduced compression thresholds. However, only a few He--RG compounds have been theoretically examined to date 
\cite{rg1,rg2}, and a comprehensive understanding of their structural and electronic behavior under extreme compression remains 
lacking. This knowledge gap primarily stems from the intrinsic complexity of these systems and the practical challenges associated 
with their experimental synthesis and characterization \cite{rg3,rg3b,rg4}.

Superionic phases, in which one atomic sublattice remains fixed while the other exhibits liquid-like diffusivity \cite{hull04}, 
have attracted great attention in recent years for their potential role in planetary interiors. In particular, first-principles 
simulations have predicted the emergence of superionic helium in He--organic mixtures (He--NH$_{3}$ and He--H$_{2}$O) at high 
pressures and temperatures, suggesting that such phases may exist deep inside the ice giants Uranus and Neptune, where they 
could significantly influence thermal transport, magnetic field generation, and the overall evolution of these planets 
\cite{superionic1,superionic2}. These predictions extend the range of planetary materials known to host superionic states, beyond 
the canonical cases of water, ammonia, or hydrogen, thus reshaping our understanding of matter under extreme conditions.

Despite this growing body of research, the possibility of superionic behavior in fully inorganic RG compounds has not been 
examined so far. Given that rare gases constitute important components in planetary interiors and atmospheres, their capacity to form 
superionic phases under compression at high temperatures could have profound implications. For instance, the stabilization of a superionic 
sublattice in He--RG systems would provide a new channel for ionic conductivity and energy transport in deep planetary layers, potentially 
altering current models of planetary structure and dynamics. Exploring such exotic phases in RG compounds is therefore not only of 
fundamental interest for condensed matter physics, but also highly relevant for Earth and planetary sciences, as it may unveil previously 
unconsidered mechanisms of heat and charge transfer in celestial bodies.

\begin{figure*}
    \centering
    \includegraphics[width=1.0\linewidth]{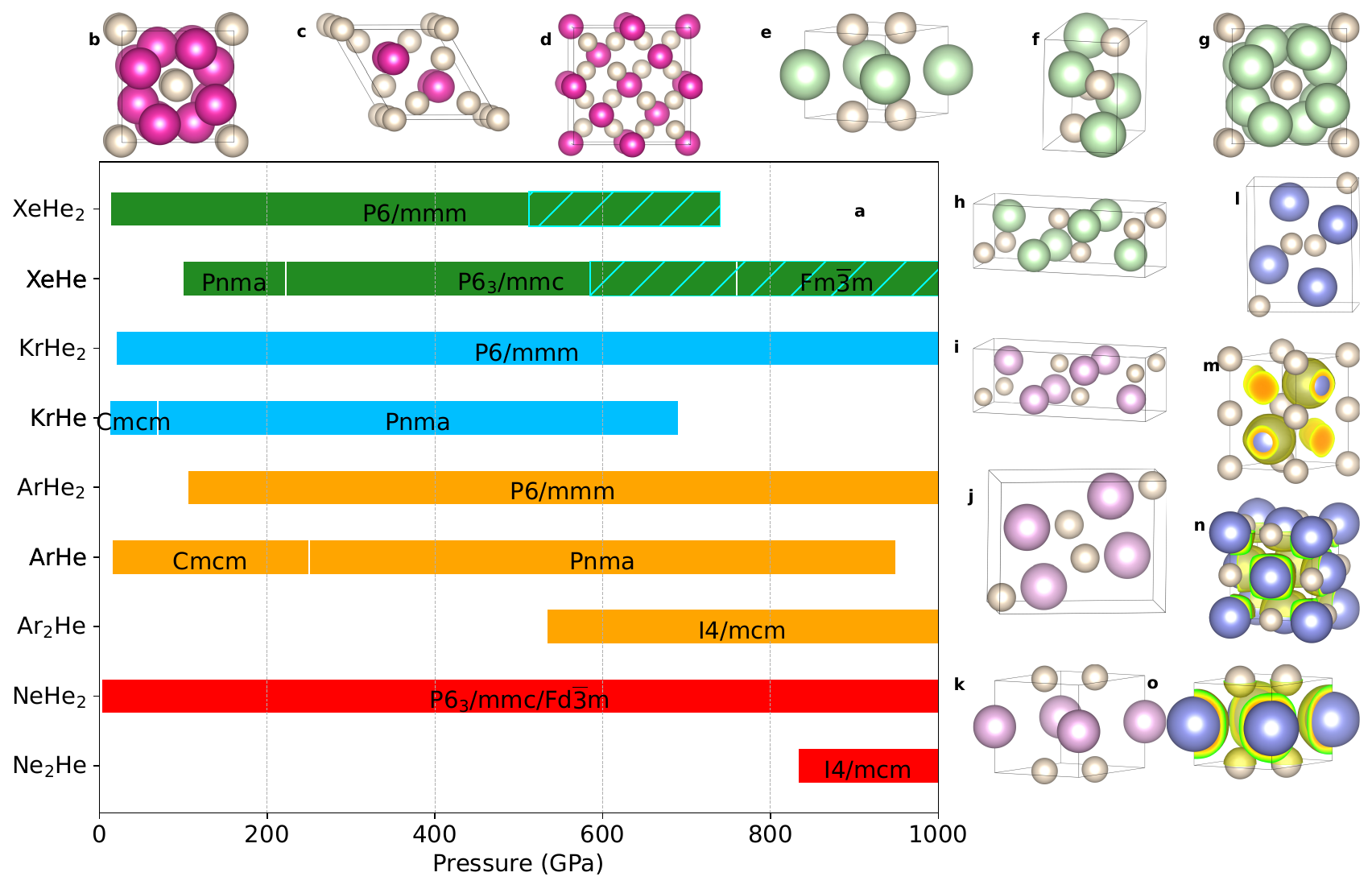}
	\caption{\textbf{Generalized zero-temperature phase diagram of He--RG compounds under pressure.}
	\textbf{a.}~Pressure--composition phase diagram of He--RG alloys; all compounds are thermodynamically stable 
	against decomposition into secondary and elemental RG phases. Cyan shaded regions denote metallic behaviour. 
	Crystal structures of \textbf{b.}~Ne$_2$He in the $I4/mcm$ phase, \textbf{c.}~NeHe$_2$ in the $P6_{3}/mmc$ phase, 
	\textbf{d.}~NeHe$_2$ in the $Fd\overline{3}m$ phase, \textbf{e.}~ArHe$_2$ in the $P6/mmm$ phase, 
	\textbf{f.}~ArHe in the $Pnma$ phase, \textbf{g.}~Ar$_2$He in the $I4/mcm$ phase, \textbf{h.}~ArHe in the
	$Cmcm$ phase, \textbf{i.}~KrHe in the $Cmcm$ phase, \textbf{j.}~KrHe in the $Pnma$ phase, \textbf{k.}~KrHe$_{2}$
	in the $P6/mmm$ phase, \textbf{l.}~XeHe in the $Pnma$ phase, \textbf{m.}~XeHe in the $P6_{3}/mmc$, 
	\textbf{n.}~XeHe in the $Fm\overline{3}m$ phase, and \textbf{o.}~XeHe$_{2}$ in the $P6/mmm$ phase.
	}
    \label{fig1}
\end{figure*}

In this work, we employed advanced crystal structure prediction techniques combined with first-principles calculations to map 
the zero-temperature phase diagram of binary He--RG compounds under compression. Our analysis reveals several previously 
unknown stoichiometric phases that are both vibrationally and thermodynamically stable against decomposition into secondary 
components, and which, importantly, can be stabilized at sub-TPa pressures, thus within the reach of current high-pressure 
experimental techniques. For He--Xe mixtures, we uncovered an even richer phenomenology, including metallic and excitonic insulator 
phases that appear at pressures dramatically lower than those required in pure helium, suggesting promising avenues for experimentally 
realizing exotic quantum states. Furthermore, we report for the first time the existence of helium-based metallic superionic phases 
in binary He--Xe compounds at elevated pressure and temperature conditions, a finding that opens new perspectives for understanding 
ionic transport in dense planetary materials. Our study not only advances fundamental knowledge on helium and RG chemistry under 
extreme conditions, but also provides a solid framework for future experimental and theoretical research in condensed matter 
physics, high-pressure chemistry, and planetary science.

\section{Results and Discussion}
\label{sec:results}
Using crystal structure prediction (CSP) methods in combination with advanced first-principles calculations based on density functional 
theory (DFT), we have determined the zero-temperature phase diagram and electronic and ionic transport properties of binary He--RG 
compounds under pressures up to $1$~TPa ($1,000$~GPa). Our computational approach included geometry optimizations, formation enthalpy 
calculations, convex hull analyses, vibrational phonon spectrum calculations, and finite-temperature \textit{ab initio} molecular dynamics 
simulations (Methods). In the following sections, we present our results for He--RG systems and discuss them, with particular emphasis 
on the newly discovered crystal structures.

\subsection{Generalized zero-temperature phase diagram}
\label{subsec:generalpd}
We restricted our analysis to binary He--RG systems with the stoichiometries AHe, A$_{2}$He, and AHe$_{2}$ (A = Ne, Ar, Kr, Xe). 
This choice is motivated by the closed-shell electronic configuration of helium ($1s^{2}$) and rare gases ($ns^{2}$$np^{6}$), 
which strongly limits their chemical reactivity under ambient conditions and suggests that, even at extreme compression, stable 
compounds are most likely to adopt simple integer ratios. These stoichiometries enable efficient atomic packing while minimizing 
Pauli repulsion and maintaining local electroneutrality, which is essential for stability in pressure-dominated regimes. By contrast, 
the formation of ternary compounds, or more complex compositions, would require substantial electronic rearrangement beyond weak 
polarization effects, making them energetically less favorable under the extreme pressure conditions considered in this study
\cite{intro2,stoi1,stoi2}.

Previous theoretical \cite{rg1,rg2,rg5} and experimental studies \cite{rg3,rg3b,rg6} have reported the stabilization of various Laves 
phases in both He--RG and RG--RG systems. Laves phases are a well-known class of structures with general formula AB$_{2}$, in which 
atoms of markedly different sizes pack efficiently into dense frameworks. They typically crystallize in one of following variants: 
cubic (C15, MgCu$_{2}$-type, space group $Fd\overline{3}m$), hexagonal (C14, MgZn$_{2}$-type, space group $P6_{3}/mmc$), or dihexagonal 
(C36, MgNi$_{2}$-type, space group $P6_{3}/mmc$). Their stability is largely governed by the atomic size mismatch, with the atomic 
radius ratio $r_{\rm A}/r_{\rm B}$ generally lying within the range $1.0$--$1.7$. This criterion is particularly relevant for He--RG 
mixtures since helium atoms are significantly smaller than other heavier RG elements \cite{rg1,rg2}. 

A non-Laves hexagonal AlB$_{2}$-type structure (space group $P6_{3}/mmc$) has also been predicted to stabilize under compression in 
AB$_{2}$ compounds with very large $r_{\rm A}/r_{\rm B}$ ratios \cite{rg1,rg2}. In particular, AHe$_{2}$ compounds have been anticipated 
to undergo the pressure-induced phase-transition sequence MgZn$_{2}$~$\to$~MgCu$_{2}$~$\to$~AlB$_{2}$ \cite{rg1}. Interestingly, the same 
sequence, but in reverse order, has been observed in Al- and Cu–rare-earth binary mixtures (e.g., ThAl$_{2}$ and YCu$_{2}$), which is 
attributed to the lower compressibility of metal–metal bonds compared to rare-earth bonds \cite{rg7,rg8,rg9}. 

By contrast, AHe and A$_{2}$He compounds have been far less investigated than the AHe$_{2}$ stoichiometry, despite indications of 
possible stability under high compression \cite{intro2,stoi1,stoi2}. To the best of our knowledge, no systematic CSP searches have 
yet been performed for AHe and A$_{2}$He. In this work, we aim to close this knowledge gap by predicting stable phases for AHe and 
A$_{2}$He systems and by characterizing their general properties, which may be of relevance for planetary science.

\begin{figure*}
    \centering
    \includegraphics[width=1.0\linewidth]{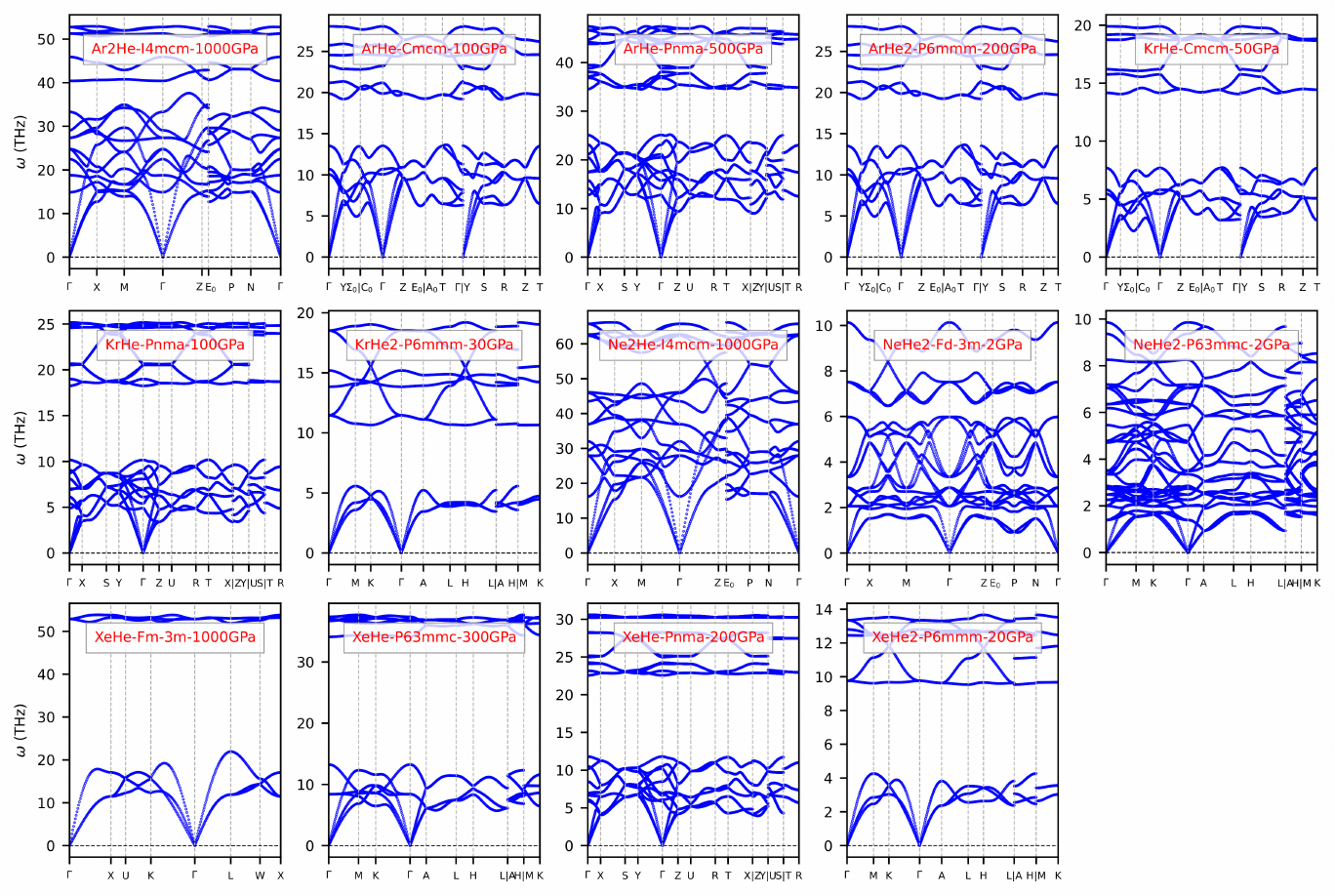}
        \caption{\textbf{Phonon spectra of highly compressed He--RG compounds.}~Phonon dispersion relations 
	are well-behaved in all cases, that is, are real and positively defined, demonstrating vibrational 
	stability of the reported crystal structures.}
    \label{fig2}
\end{figure*}

Next, we present the formation enthalpy results obtained for each He--RG system and identify the phases that are likely to be stable 
at zero temperature as a function of pressure up to $1,000$~GPa. It is important to emphasize that all phases displayed in Fig.~\ref{fig1}a 
are thermodynamically stable against decomposition into secondary and/or elemental phases, as verified by convex-hull calculations 
(not shown in the main text). Furthermore, these phases are also vibrationally stable, as confirmed by the phonon dispersion relations 
in Fig.~\ref{fig2}, which exhibit no imaginary frequencies. We conclude this section by discussing the general zero-temperature 
phase-diagram trends across He--RG systems, with particular attention to the role of RG atomic radius.
\\

\begin{figure*}
    \centering
    \includegraphics[width=1.0\linewidth]{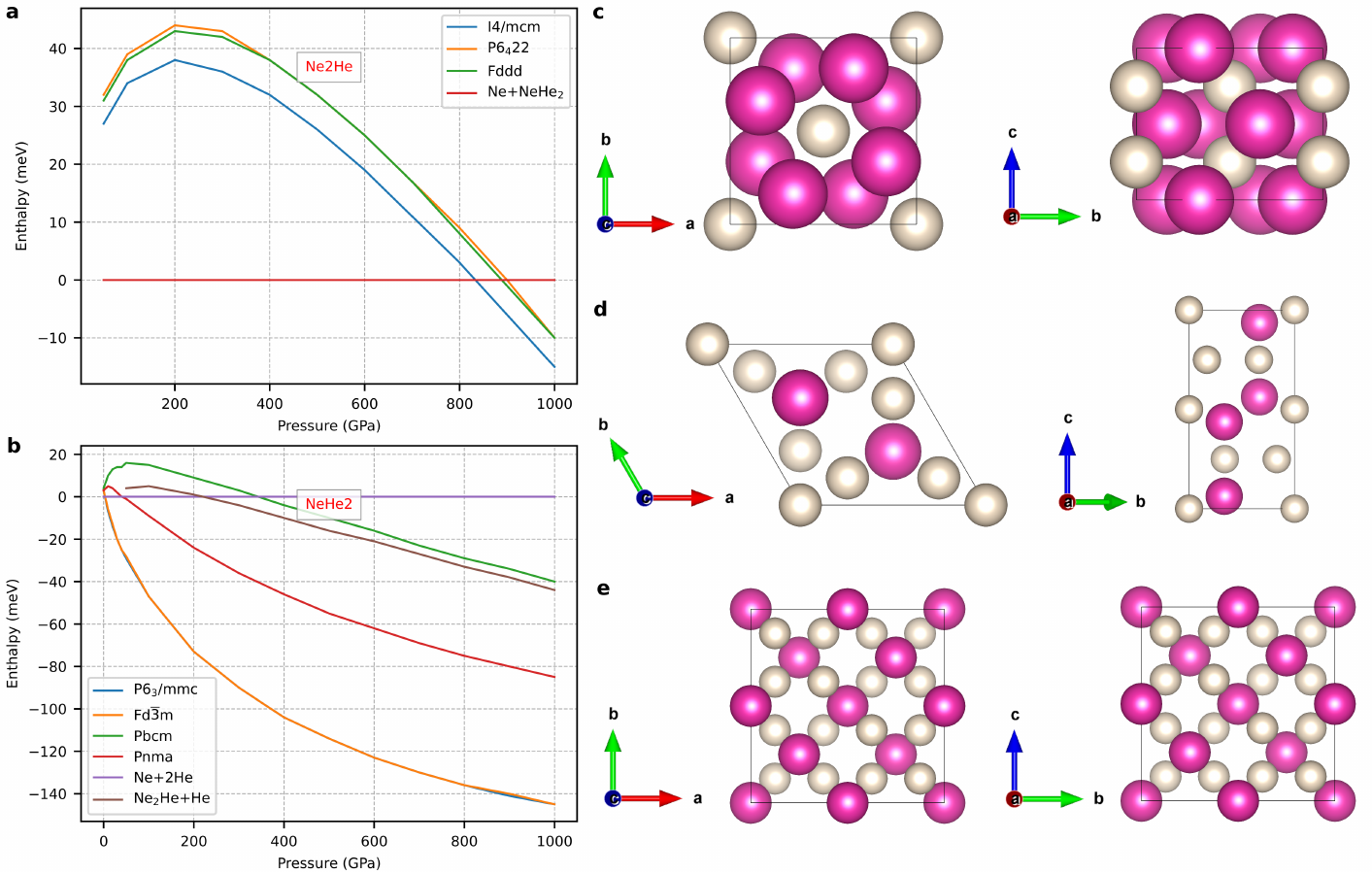}
        \caption{\textbf{Formation enthalpy of He--Ne compounds under pressure.}~The NeHe composition, not shown
	in the figure, is thermodynamically unstable against elemental decomposition into Ne and He.
	\textbf{a.}~Ne$_{2}$He composition.
	\textbf{b.}~NeHe$_{2}$ composition. Ball-stick representation of the
	\textbf{c.}~tetragonal $I4/mcm$, 
	\textbf{d.}~hexagonal $P6_{3}/mmc$, and
	\textbf{e.}~cubic $Fd\overline{3}m$ phases. 
	Ne and He atoms are represented by magenta and white spheres, respectively.
        }
    \label{fig3}
\end{figure*}

\textit{He--Ne compounds.}~Figure~\ref{fig3} shows the formation enthalpy curves of He--Ne systems at zero temperature as a 
function of pressure. The NeHe$_{2}$ and Ne$_{2}$He stoichiometries are found to form thermodynamically stable compounds at 
approximately $0$ and $835$~GPa, respectively. In contrast, the NeHe stoichiometry, not shown in Fig.~\ref{fig1}, remains 
unstable with respect to decomposition into He and Ne across the investigated pressure range.

According to our CSP searches and DFT calculations, Ne$_{2}$He stabilizes under very high compression in a centrosymmetric tetragonal 
phase with $I4/mcm$ symmetry (Figs.~\ref{fig3}a,c). The corresponding unit cell contains four formula units ($Z = 4$), with the $c$ 
lattice parameter significantly shorter than the two equivalent $a$ and $b$ parameters (Supplementary Fig.~1). Viewed along the $c$ 
axis, the Ne sublattice adopts a spiral-like arrangement (Fig.~\ref{fig3}c). Other cubic and hexagonal structures identified in our 
CSP searches are found to be energetically competitive with the predicted tetragonal ground-state phase, although remain metastable 
across the entire analysed pressure range (Fig.~\ref{fig3}a).

Regarding NeHe$_{2}$ (Fig.~\ref{fig3}b), the enthalpies of the Laves hexagonal $P6_{3}/mmc$ ($Z = 4$) and cubic $Fd\overline{3}m$ 
($Z = 8$) phases (Figs.~\ref{fig3}d,e) remain practically indistinguishable within our numerical accuracy of $1$~meV per formula unit 
(Supplementary Figs.~2--3 and Supplementary Discussion). These theoretical results are consistent with low-pressure experimental observations 
that assign the atomic structure of NeHe$_{2}$ to the hexagonal Laves phase MgZn$_{2}$ \cite{rg3b}. Similarly, our calculations agree 
with previous DFT studies performed with functionals not accounting for long-range van der Waals interactions, which predicted a 
pressure-induced MgZn$_{2}$~$\to$~MgCu$_{2}$ phase transition at approximately $120$~GPa \cite{rg1}.
\\

\begin{figure*}
    \centering
    \includegraphics[width=0.9\linewidth]{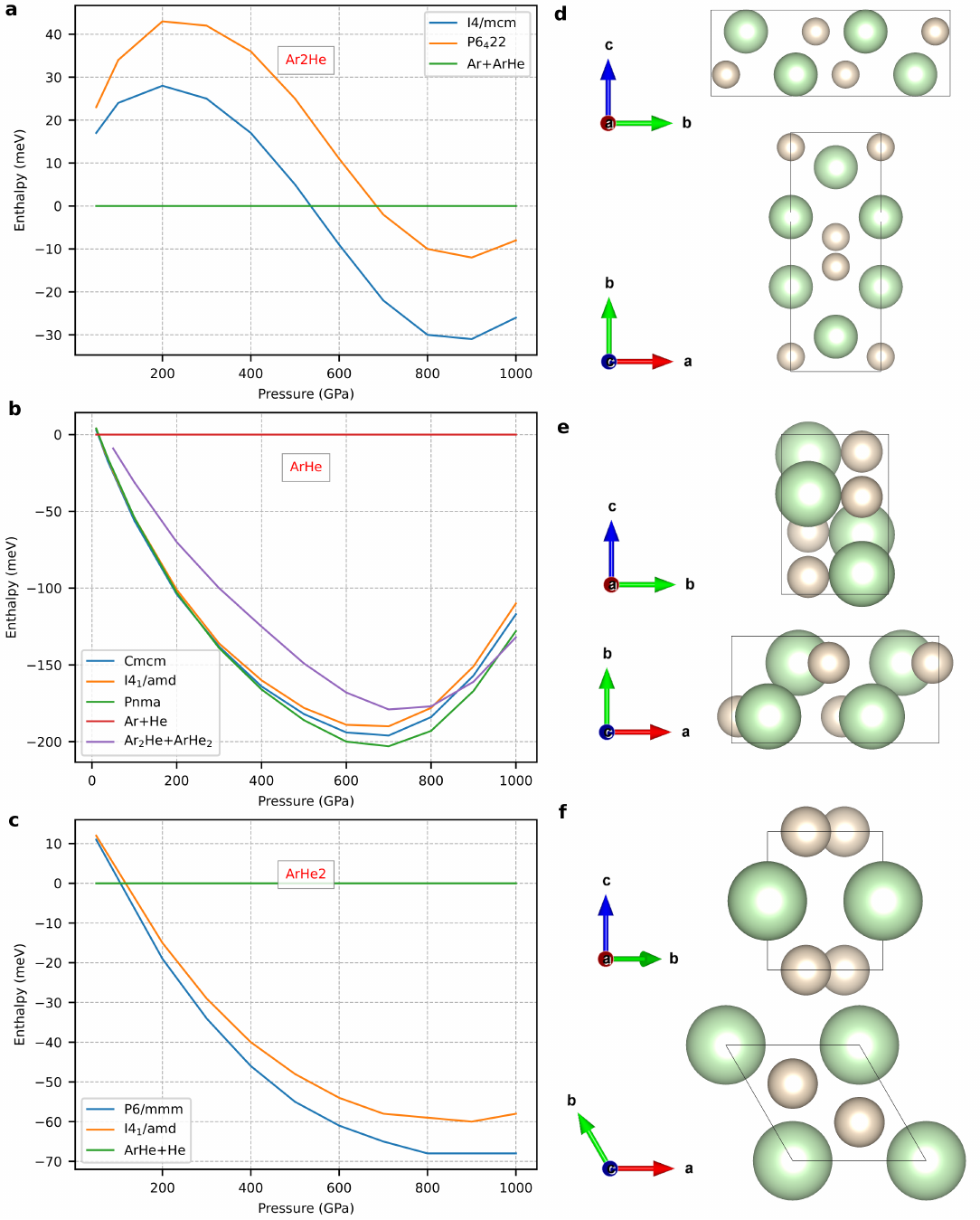}
 \caption{\textbf{Formation enthalpy of He--Ar compounds under pressure.}
        \textbf{a.}~Ar$_{2}$He composition.
        \textbf{b.}~ArHe$_{2}$ composition. 
        \textbf{c.}~ArHe$_{2}$ composition. Ball-stick representation of the
        \textbf{d.}~orthorhombic $Cmcm$,
        \textbf{e.}~orthorhombic $Pnma$, and
        \textbf{f.}~hexagonal $P6/mmm$ phases.
        Ar and He atoms are represented by green and white spheres, respectively.
        }
	\label{fig4}
\end{figure*}

\textit{He--Ar compounds.}~The Ar$_{2}$He system is predicted to behave very similarly to Ne$_{2}$He (Fig.~\ref{fig1}a). In particular, 
its lowest-enthalpy phase is a centrosymmetric tetragonal structure with $I4/mcm$ symmetry (Fig.~\ref{fig4}a) and $Z = 4$ (Supplementary 
Fig.~4). The main difference compared with the analogous Ne-based system is that the threshold pressure for stabilization against 
decomposition into secondary phases is significantly lower, at about $520$~GPa. An enantiomorphic hexagonal phase with space group 
$P6_{4}22$ is found to be energetically competitive with the predicted ground-state phase; however, it remains consistently higher in 
energy by several tens of meV per formula unit across the entire analysed pressure range (Fig.~\ref{fig4}a).

Ar--He compounds with 1:1 composition ratio are predicted to be thermodynamically stable from relatively low pressures (a few 
tens of GPa) up to $950$~GPa (Fig.~\ref{fig4}b). Under low compression, the lowest-enthalpy structure of ArHe is found to be a 
centrosymmetric orthorhombic phase with space group $Cmcm$ (Figs.~\ref{fig4}b,d). The corresponding unit cell contains four formula 
units ($Z = 4$) and is markedly elongated along one lattice vector (Supplementary Fig.~5). At around $250$~GPa, this orthorhombic 
$Cmcm$ phase transforms into another orthorhombic structure with space group $Pnma$ (Figs.~\ref{fig4}b,e). This high-pressure phase 
also contains eight atoms per unit cell, although its lattice vectors are more similar in length than in the low-pressure phase 
(Supplementary Fig.~6). To the best of our knowledge, neither of these two orthorhombic phases has been previously predicted for 
ArHe or for any other RG--RG compound with 1:1 composition ratio.

Regarding ArHe$_{2}$, we find a completely different phase competition scenario compared with NeHe$_{2}$ (Fig.~\ref{fig4}c). 
Specifically, no Laves phase emerges as the ground state under pressure; instead, a new hexagonal phase with space group $P6/mmm$ 
becomes the most favorable structure within the analyzed pressure range. This phase is predicted to be thermodynamically stable 
against decomposition into secondary phases at pressures above $\approx 100$~GPa. The previously overlooked $P6/mmm$ phase is highly 
symmetric and contains only three atoms per unit cell ($Z = 1$, Fig.~\ref{fig4}f and Supplementary Fig.~7). Notably, its enthalpy 
lies well below those of the well-known $Fd\overline{3}m$ and $P6_{3}/mmc$ Laves phases, regardless of the employed DFT 
exchange–correlation functional (Supplementary Discussion). We also note that a centrosymmetric tetragonal phase with $I4_{1}/amd$ 
symmetry is found to be close in energy to the predicted hexagonal ground state (Fig.~\ref{fig4}c).
\\

\begin{figure*}
    \centering
    \includegraphics[width=1.0\linewidth]{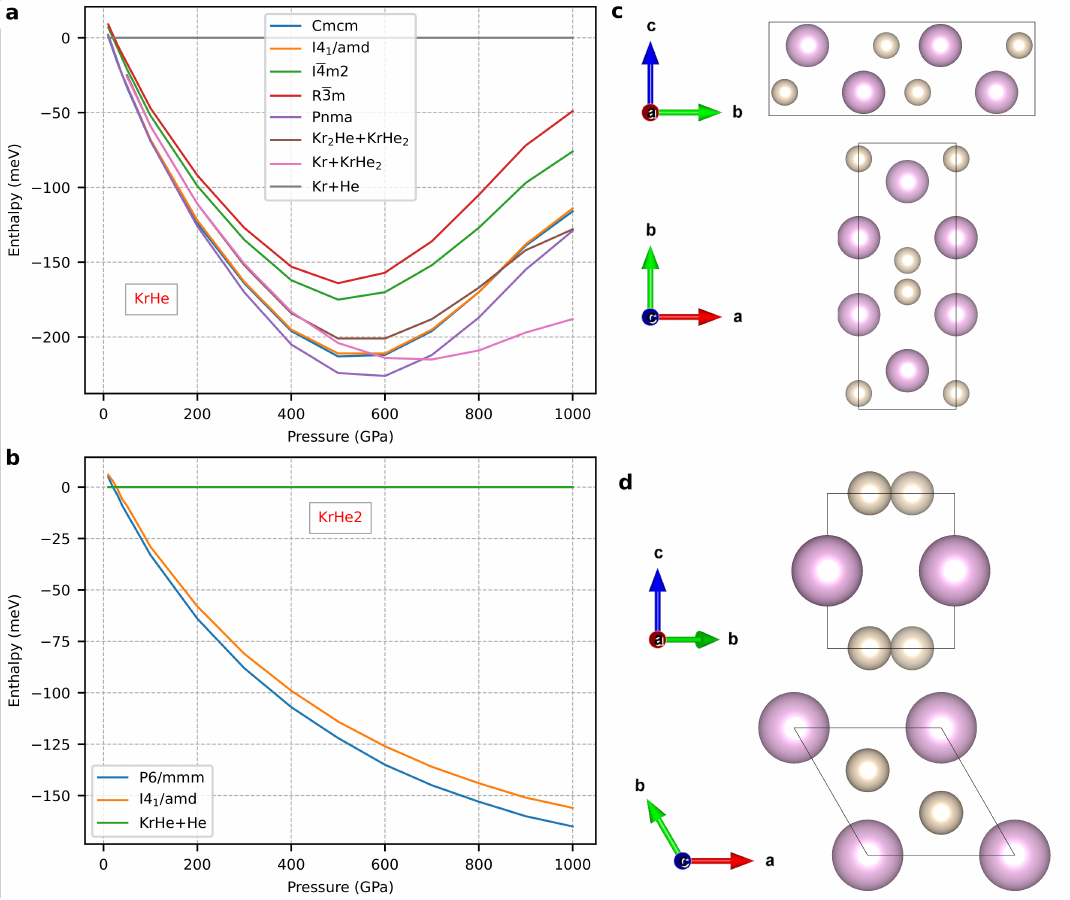}
	\caption{\textbf{Formation enthalpy of He--Kr compounds under pressure.}~The Kr$_{2}$He composition, not 
	shown in the figure, is thermodynamically unstable against elemental decomposition into Kr and He.
	\textbf{a.}~KrHe composition.
        \textbf{b.}~KrHe$_{2}$ composition. Ball-stick representation of the
        \textbf{c.}~orthorhombic $Cmcm$, and
        \textbf{d.}~hexagonal $P6/mmm$ phases.
        Kr and He atoms are represented by violet and white spheres, respectively.
        }
	\label{fig5}
\end{figure*}

\textit{He--Kr compounds.}~No phase was found to be thermodynamically stable for compounds with the Kr$_{2}$He stoichiometry. 
In contrast, KrHe is predicted to exhibit a stability window in the pressure range $20 \le P \le 680$~GPa, within which a 
$Cmcm$~$\to$~$Pnma$ phase transition occurs at approximately $80$~GPa (Fig.~\ref{fig5}a). The two orthorhombic structures 
involved in this pressure-induced transformation are equivalent to those described above for ArHe (Fig.~\ref{fig5}c). In 
addition, a centrosymmetric tetragonal structure with $I4/mcm$ symmetry is found to be energetically competitive with the 
predicted ground-state phase (Fig.~\ref{fig5}a).

For KrHe$_{2}$, the ground-state scenario closely parallels that of ArHe$_{2}$. In particular, a highly symmetric 
hexagonal $P6/mmm$ phase emerges as the lowest-enthalpy structure (Figs.~\ref{fig5}b,d). The main difference between 
KrHe$_{2}$ and ArHe$_{2}$ lies in their thermodynamic stability ranges: in KrHe$_{2}$, the hexagonal $P6/mmm$ phase 
persists over a much broader pressure interval, from approximately $10$~GPa up to $1,000$~GPa. Interestingly, the 
phonon dispersion relation calculated for this phase (Fig.~\ref{fig2}) reveals a wide frequency gap between the 
low-energy acoustic and high-energy optical branches, a vibrational feature characteristic of layered structures.
\\

\begin{figure*}
    \centering
    \includegraphics[width=1.0\linewidth]{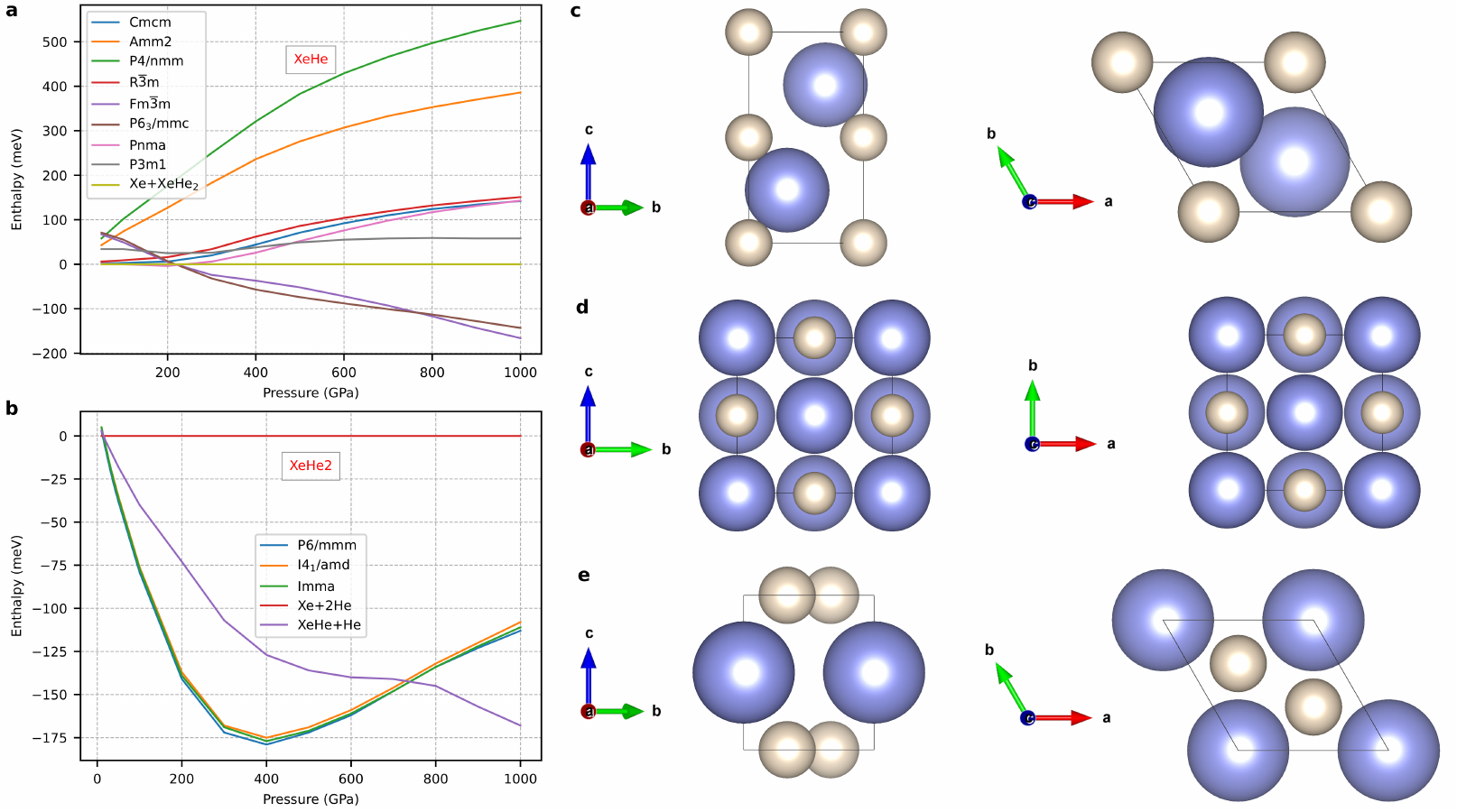}
	\caption{\textbf{Formation enthalpy of He--Xe compounds under pressure.}~The Xe$_{2}$He composition, not shown
        in the figure, is thermodynamically unstable against elemental decomposition into Xe and He.
        \textbf{a.}~XeHe composition.
        \textbf{b.}~XeHe$_{2}$ composition. Ball-stick representation of the
        \textbf{c.}~hexagonal $P6_{3}/mmc$,
	\textbf{d.}~cubic $Fm\overline{3}m$, and
	\textbf{e.}~hexagonal $P6/mmm$ phases.
        Xe and He atoms are represented by blue and white spheres, respectively.
        }
	\label{fig6}
\end{figure*}

\textit{He--Xe compounds.}~Similarly to the previous case, no thermodynamically stable phase was identified across the entire 
pressure range for compounds with the Xe$_{2}$He stoichiometry. In contrast, XeHe exhibits a sequence of pressure-induced phase 
transitions: an orthorhombic $Pnma$ phase, stabilized at about $100$~GPa, transforms into a hexagonal $P6_{3}/mmc$ phase at 
approximately $220$~GPa, which in turn transitions into a cubic $Fm\overline{3}m$ phase at $770$~GPa (Fig.~\ref{fig6}a). The 
previously unknown hexagonal $P6_{3}/mmc$ phase contains four atoms per unit cell ($Z = 2$, Fig.~\ref{fig6}c) and is characterized 
by a markedly elongated $c$ axis (Supplementary Fig.~8). The also newly identified cubic $Fm\overline{3}m$ phase contains eight 
atoms per unit cell ($Z = 4$, Fig.~\ref{fig6}d) and achieves highly efficient atomic packing (Supplementary Fig.~9).

For XeHe$_{2}$, the ground state is also a highly symmetric hexagonal $P6/mmm$ phase (Figs.~\ref{fig6}b,e), structurally analogous 
to the phases predicted for KrHe$_{2}$ and ArHe$_{2}$. This phase becomes stabilized at pressures above $\sim 20$~GPa but decomposes 
into a mixture of XeHe and He at around $750$~GPa (Fig.~\ref{fig6}b). Notably, in contrast to XeHe$_{2}$, the same hexagonal phase 
in KrHe$_{2}$ and ArHe$_{2}$ remains stable against decomposition even at the highest pressures considered (Fig.~\ref{fig1}a).
\\

\begin{figure*}
    \centering
    \includegraphics[width=0.8\linewidth]{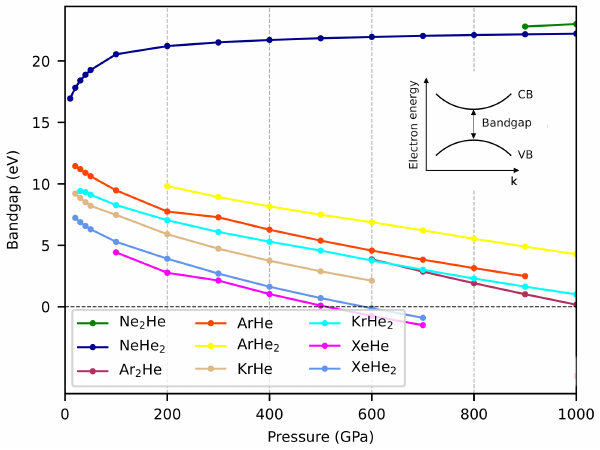}
        \caption{\textbf{Pressure dependence of the bandgap in thermodynamically and vibrationally stable 
        He--RG compounds.}~Metallic behaviour emerges when the bandgap, defined as the energy difference
        between the bottom of the conduction band (CB) and top of the valence band (VB), is zero or negative.
        }
    \label{fig7}
\end{figure*}

\textit{General phase-diagram trends.}~The stability of He--RG compounds strongly depends on both the stoichiometry and atomic radius 
of the RG species. For the smallest RG, neon, only the NeHe$_{2}$ and Ne$_{2}$He compositions form stable compounds, while the 1:1 
stoichiometry remains unstable across all pressures. Moving to larger atoms (Ar, Kr, Xe), the range of stable stoichiometries broadens: 
ArHe and KrHe are stable over wide pressure windows, undergoing $Cmcm$~$\to$~$Pnma$ transitions, while XeHe exhibits an even richer 
sequence of orthorhombic, hexagonal, and cubic phases with increasing pressure. In contrast, the A$_{2}$He stoichiometry is progressively 
destabilized with increasing rare-gas size: while Ne$_{2}$He and Ar$_{2}$He form stable $I4/mcm$ tetragonal phases, Kr$_{2}$He and 
Xe$_{2}$He show no thermodynamically stable phases. Conversely, the AHe$_{2}$ stoichiometry becomes increasingly favored with larger 
RG atoms, stabilizing in a simple hexagonal $P6/mmm$ lattice for ArHe$_{2}$, KrHe$_{2}$, and XeHe$_{2}$, persisting over very broad 
pressure intervals.

In terms of crystal symmetry, the well-known Laves phases (cubic $Fd\overline{3}m$ and hexagonal $P6_{3}/mmc$) dominate the structural 
landscape of NeHe$_{2}$, but they are replaced by the new high-symmetry $P6/mmm$ phase in ArHe$_{2}$, KrHe$_{2}$, and XeHe$_{2}$. The 
cubic $Fm\overline{3}m$ phase also emerges uniquely in XeHe at very high compression. Overall, a clear trend arises: smaller RG atoms 
stabilize more complex and distorted structures (e.g., Ne$_{2}$He tetragonal, Laves-type NeHe$_{2}$), whereas heavier RG atoms promote 
simple high-symmetry hexagonal and cubic packings. This distinction highlights a fundamental structural difference between Ne--He 
compounds and the heavier He--RG systems: Ne--He mixtures favor Laves-type geometries with competing near-degenerate polymorphs, while 
the larger He--RG compounds adopt the simpler $P6/mmm$ hexagonal phase that remains robust across extensive pressure ranges.
\\

\subsection{Electronic properties}
\label{subsec:electronic} 
After establishing the generalized phase diagram of highly compressed He--RG systems, we turn our attention to their electronic 
properties. We start focusing on the pressure dependence of the electronic band gap, $E_{g}$ (Fig.~\ref{fig7}). This electronic 
band gap, defined as the energy difference between the highest occupied electronic states (top of the valence band, VB) and the 
lowest unoccupied states (bottom of the conduction band, CB), governs the ease with which the material conducts electricity and 
absorbs light. To accurately capture electronic correlations and spatial localization effects, we employed a hybrid DFT 
exchange--correlation functional \cite{hse06} for this part of our study.

Figure~\ref{fig7} shows the pressure dependence of the electronic band gap for He–RG systems, neglecting possible quantum nuclear and 
thermal effects \cite{cazorla17,benitez25a,benitez25b}. The results are presented for the nine thermodynamically stable compounds identified 
in the previous section. A clear general trend emerges: in He--Ne systems, increasing pressure widens the band gap, thereby reinforcing 
their insulating character, whereas in the remaining He--RG systems pressure progressively narrows the band gap, ultimately driving them 
towards metallicity. In particular, XeHe and XeHe$_{2}$ are predicted to become metallic at sub-terapascal pressures of approximately 
$510$ and $585$~GPa, respectively. 

At the highest pressure considered in this work, Ar$_{2}$He is also found to undergo an insulator-metal transition. Furthermore, by 
extrapolating the $E_{g}$ curves in Fig.~\ref{fig7}, we anticipate that the other Ar-- and Kr--based compounds (i.e., ArHe, ArHe$_{2}$, 
KrHe, and KrHe$_{2}$) will likewise become metallic at pressures not far above $1$~TPa. Remarkably, these transition pressures 
are much lower than the critical pressure estimated for pure $^{4}$He, set at around $25$~TPa \cite{cong23,dmc1,dmc2}. Thus, alloying helium 
with heavier rare gases emerges as an efficient strategy to induce metallicity in He-rich compounds under conditions that could, in principle,
be realized in high-pressure laboratory experiments on Earth \cite{lab1,lab2,lab3}.

\begin{figure*}
    \centering
    \includegraphics[width=1.0\linewidth]{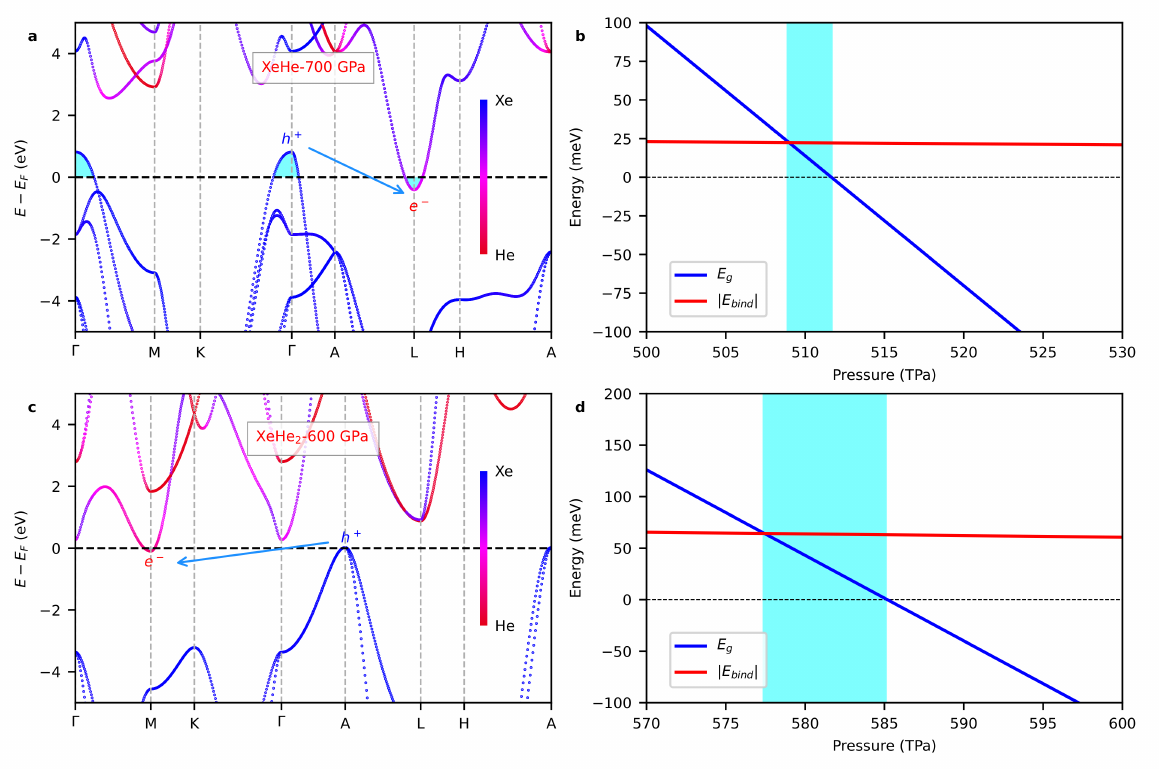}
        \caption{\textbf{Electronic properties of He--Xe compounds under pressure.}~Band structure of metallic
	\textbf{a.}~XeHe in the hexagonal $P6_{3}/mmc$ phase, and
	\textbf{b.}~XeHe$_{2}$ in the hexagonal $P6/mmm$ phase. 
	Comparison of the excitonic binding energy, $E_{\rm bind}$, and band gap, $E_{g}$, in \textbf{c.}~XeHe 
	and \textbf{d.}~XeHe$_{2}$ as a function of pressure. Cyan shaded regions indicate the pressure intervals 
	in which the sufficient condition for spontaneous formation of excitons, $E_{g} \le |E_{\rm bind}|$, is fulfilled.
	}
    \label{fig8}
\end{figure*}

\begin{figure*}
    \centering
    \includegraphics[width=1.0\linewidth]{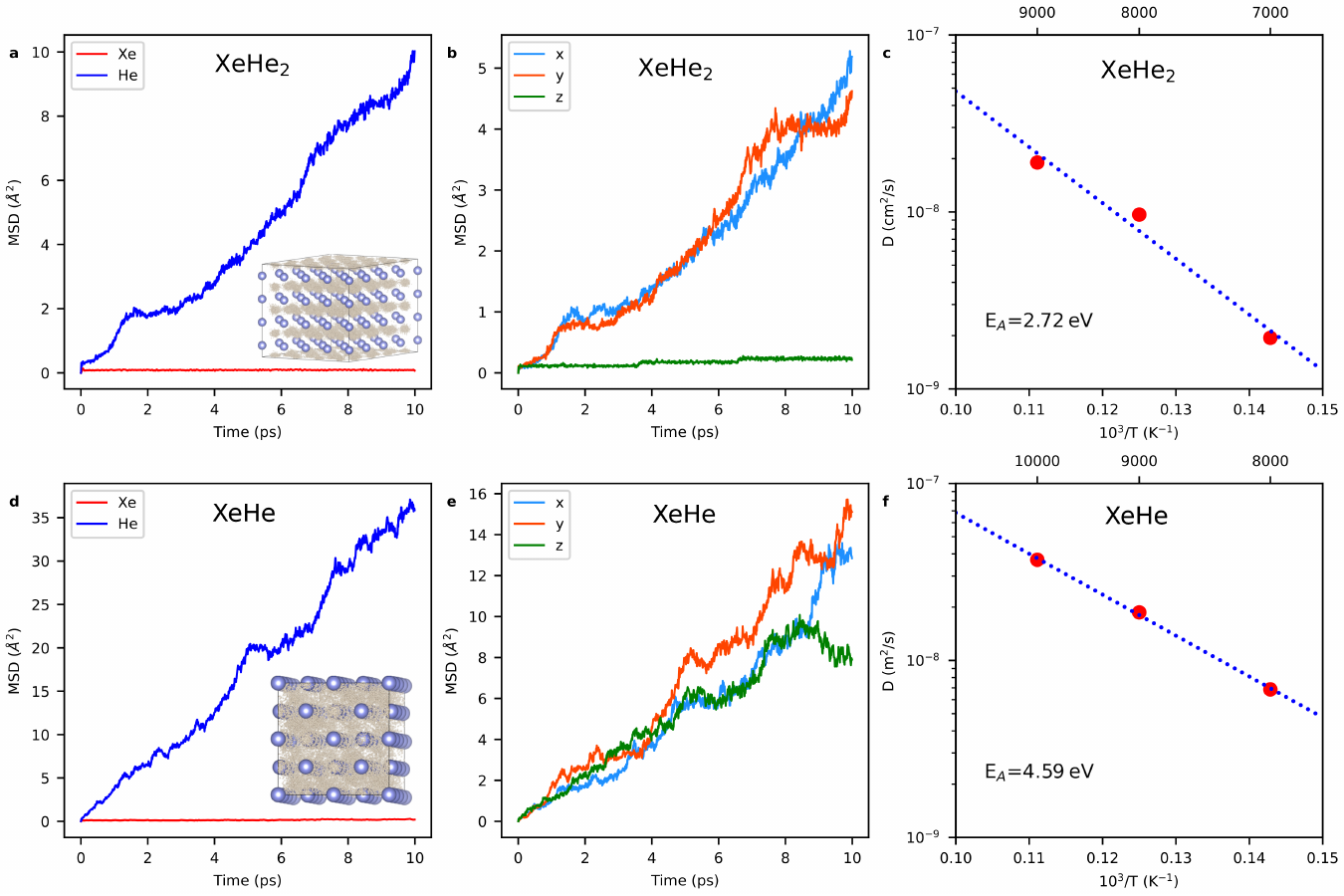}
        \caption{\textbf{Ionic diffusion properties of He--Xe compounds under extreme pressure and temperature
	conditions.}~Atomic mean squared displacement (MSD) estimated from AIMD simulations for 
	\textbf{a.}~XeHe$_{2}$ in the hexagonal $P6/mmm$ phase at $600$~GPa and $8,000$~K, and
	\textbf{d.}~XeHe in the cubic $Fm\overline{3}m$ phase at $1,000$~GPa and $10,000$~K. 
	In superionic XeHe$_{2}$, helium ions diffuse anisotropically within the x--y plane, as shown in \textbf{b.} 
	and the inset of \textbf{a.} (where several atomic trajectories are represented with points). 
        In superionic XeHe, the helium atoms diffuse isotropically through the entire volume, as shown in \textbf{e.} 
	and the inset of \textbf{d.} (where several atomic trajectories are represented with points).
        In superionic XeHe$_{2}$ and XeHe, the activation energy for helium diffusion amounts to 
	\textbf{c.}~$2.7$~eV at $600$~GPa and \textbf{f.}~$4.6$~eV at $1,000$~GPa, respectively.
	}
    \label{fig9}
\end{figure*}

The pressure derivative of the band gap, $\alpha$, is found to be nearly identical across all He--RG compounds, with the exception 
of Ne$_{2}$He and NeHe$_{2}$ (Fig.~\ref{fig7}). Specifically, $\alpha$ adopts a value of approximately $-12$~meV~GPa$^{-1}$ in all 
compounds exhibiting band-gap closure. This observation suggests that metallicity in He--RG systems is governed primarily by 
composition rather than by structural features. Indeed, although Ne$_{2}$He and Ar$_{2}$He adopt the same crystal structure at high 
pressures (Fig.~\ref{fig1}a), $\alpha$ is positive in the former while negative in the latter.

Motivated by recent theoretical predictions that $^{4}$He may become an excitonic insulator (EI) in the TPa regime \cite{cong23}, we
extended our electronic analysis of He--RG systems. An EI is a phase of matter that can arise in narrow-band gap semiconductors or semimetals 
when the Coulomb attraction between electrons and holes is sufficiently strong to induce the spontaneous condensation of bound electron–hole 
pairs (i.e., excitons), even without external excitation. In this state, the exciton binding energy $E_{\rm bind}$ is negative and larger in
magnitude than the band gap ($E_{g} \le |E_{\rm bind}|$), leading to the formation of a macroscopic exciton condensate in the ground state.
Since this condition can be met near band-gap closure, we examined in detail the electronic band structures of XeHe and XeHe$_{2}$, that
is, the two He--RG compounds predicted to undergo metallization below $1$~TPa.

Figures~\ref{fig8}a,c display the electronic band structures of XeHe and XeHe$_{2}$ calculated near their insulator-to-metal transition
pressure. In both cases, the systems exhibit an indirect band gap since the valence-band maximum and conduction-band minimum are located 
at different reciprocal space points. Like in solid helium, the small overlap between the conduction and valence bands yields a very low 
electronic density of states at the Fermi level, characteristic of semimetallic behaviour. Figures~\ref{fig8}b,d compare the values of 
$E_{\rm bind}$ and $E_{g}$ calculated for the two compounds under similar pressure conditions. The exciton binding energies were 
evaluated using the Wannier-Mott formula and a first-principles approach equivalent to that employed in Ref.~\cite{cong23}, yielding 
$|E_{\rm bind}| = 25$~meV for XeHe and $60$~meV for XeHe$_{2}$ (Figs.~\ref{fig8}b,d). 

Our results indicate that XeHe hosts an EI phase at around $510$~GPa, which is stable within a narrow pressure window of a few GPa. In 
contrast, XeHe$_{2}$, for which the calculated exciton binding energy is much larger, develops an EI state at approximately $580$~GPa 
that persists over a broader pressure interval of approximately $8$~GPa. These critical pressures are nearly an order of magnitude lower 
than those predicted for pure $^{4}$He. Hence, mixing helium with heavier RG appears to be an effective route to stabilize exotic states 
of matter such as the EI phase in He-rich compounds under conditions attainable in high-pressure Earth laboratories.

\subsection{Ionic transport properties}
\label{subsec:superionic}
The possibility of stabilizing superionic helium in He--organic mixtures at high pressures and temperatures has recently attracted 
considerable attention in planetary science \cite{superionic1,superionic2}. Superionic helium provides a pathway to unique transport 
properties and phase behaviours, like enhanced miscibility and unexpected chemical reactivity, that may shed light on long-standing 
problems such as the helium depletion in the atmospheres of Jupiter and Saturn, as well as the anomalous magnetic fields and heat 
fluxes of Uranus and Neptune \cite{puzzle1,puzzle2}. By contrast, the possibility of helium superionicity in He--inorganic mixtures 
(i.e., systems not containing C, H, O, or N atoms) has been far less explored, despite its potential relevance for understanding the 
interiors of giant planets and improving current models of planetary formation \cite{puzzle3}. This knowledge gap partly motivates the 
present study, in which we investigate the superionic behaviour of helium in He--RG compounds.

Figure~\ref{fig9} presents the results of our \textit{ab initio} molecular dynamics (AIMD) simulations for metallic XeHe and XeHe$_{2}$ 
at high pressures and temperatures. Both compounds exhibit clear superionic behaviour under the selected conditions, as evidenced by 
the mean squared displacement (MSD) plots in Figs.~\ref{fig9}a,d, which show a nonzero (null) slope for helium (RG) atoms. These compounds 
were chosen because the coexistence of metallicity and helium diffusion is particularly intriguing. Due to the substantial computational 
cost of AIMD simulations, we were able to study only these representative cases. Nonetheless, the possibility that other He--RG compounds 
may also exhibit superionicity under extreme pressure and temperature conditions cannot be ruled out.

Our AIMD simulations performed at fixed pressure and temperature reveal strongly anisotropic helium diffusion in XeHe$_{2}$ 
(Fig.~\ref{fig9}b), in contrast to the fully isotropic behaviour observed in XeHe (Fig.~\ref{fig9}e). This difference in 
ionic transport arises from their distinct crystal structures: hexagonal and layered-like ($P6/mmm$) in XeHe$_{2}$ versus cubic and 
isotropic ($Fm\overline{3}m$) in XeHe. The energy barriers for helium diffusion, extracted from the temperature dependence of the 
MSD curves (Figs.~\ref{fig9}c,f), are $2.72$~eV in XeHe$_{2}$ at $600$~GPa and $4.59$~eV in XeHe at $1,000$~GPa. Although these values 
are not directly comparable due to the different pressure conditions, their magnitude indicates that extremely high temperatures, of 
the order of $10,000$~K, are required to activate helium transport in both compounds. Notably, the two crystal structures in which 
helium superionicity emerges had not been reported before and are predicted in this study for the first time in He–Xe systems.

To the best of our knowledge, this is the first prediction of a helium phase that is both superionic and metallic in a mixture of 
inorganic compounds. The implications of this theoretical finding are far-reaching. If helium were to exhibit superionic behaviour 
deep inside giant planets (e.g., Jupiter, Saturn, or exoplanets subject to even more extreme pressure–temperature conditions), it 
would profoundly affect our understanding of thermal conductivity and mass transport in their interiors. The existence of a metallic 
helium-rich phase would further imply that helium, typically regarded as inert and insulating, could contribute to planetary dynamos 
beyond hydrogen metallic layers, potentially accounting for anomalies in observed magnetic fields. Moreover, the impact on the equation 
of state of giant planets would be significant, as current models generally treat helium as insulating and immiscible. Finally, the 
stabilization of a helium–xenon metallic and superionic phase could offer a mechanism for noble gas sequestration or unusual 
high-pressure solubility, thereby helping to resolve long-standing puzzles regarding noble gas abundances in gas giant atmospheres.

\section{Conclusions}
\label{sec:conclusions}
Our comprehensive first-principles study establishes the existence of several previously unknown helium--RG compounds that are stable 
at sub-TPa pressures. Across different stoichiometries, we uncovered a series of high-symmetry phases, most notably a $P6/mmm$ 
hexagonal structure for ArHe$_{2}$, KrHe$_{2}$, and XeHe$_{2}$, that persists over broad pressure ranges. Likewise, the simple binary 
compounds ArHe, KrHe and XeHe are found to adopt thermodynamically stable orthorhombic and cubic phases. Quantitatively, we find 
metallization thresholds of approximately $510$~GPa for XeHe and $585$~GPa for XeHe$_{2}$, nearly an order of magnitude lower than the 
critical value predicted for pure helium ($\approx 25$~TPa). Moreover, excitonic insulating states emerge in these xenon-rich compounds 
within narrow but experimentally accessible pressure windows, providing an efficient route to stabilize exotic quantum condensates.

Finite-temperature simulations reveal the onset of helium superionicity in XeHe and XeHe$_{2}$ under extreme pressure--temperature 
conditions. In XeHe$_{2}$, helium diffusion proceeds anisotropically within layered hexagonal planes, while in cubic XeHe it occurs 
isotropically. The estimated diffusion barriers point to activation temperatures of $\sim 10,000$~K, consistent with conditions found 
in giant planetary interiors. These results represent the first theoretical prediction of metallic and superionic helium phases in 
inorganic mixtures, highlighting a unique regime where electronic and ionic transport coexist.

The implications of our findings extend well beyond high-pressure chemistry. In planetary science, they suggest new mechanisms for 
energy and charge transport in the deep interiors of gas giants and exoplanets, potentially contributing to unexplained anomalies in 
magnetic field generation and thermal evolution. In condensed matter physics, they demonstrate that alloying helium with heavier RG 
is a practical strategy to access exotic quantum states, such as metallicity, excitonic insulators, and superionicity, within 
experimentally achievable pressure ranges. This work thus establishes a foundation for future experimental exploration and opens 
transformative perspectives across high-pressure physics and planetary science.

\section*{Methods}
\label{sec:methods}
{\bf First-principles simulations.}~\textit{Ab initio} calculations based on density functional theory (DFT) 
\cite{cazorla17} were performed to analyse the structural, phase stability and electronic properties of He--RG compounds. 
We performed these calculations with the VASP code \cite{vasp} using the PBEsol \cite{pbesol} approximation to the
exchange-correlation energy. Dispersion van der Waals interactions were accounted for within the optB88 framework 
\cite{optb88}. The projector augmented-wave method was used to represent the ionic cores \cite{bloch94} and the following 
electronic states were considered as valence: He $1s$, Ne $2s$ $2p$, Ar $3s$ $3p$, Kr $4s$ $4p$, and Xe $5s$ $5p$. 
Wave functions were represented in a plane-wave basis typically truncated at $650$~eV. By using these parameters and 
dense {\bf k}-point grids for reciprocal-space integration, zero-temperature energies were converged to within $0.5$~meV 
per formula unit. In the geometry relaxations, a force tolerance of $0.005$~eV$\cdot$\AA$^{-1}$ was imposed in all the 
atoms.
\\

{\bf Phonons calculations.}~The second-order interatomic force constant of all He--RG compounds and the resulting 
harmonic phonon spectrum were calculated with the finite-differences method as is implemented in the \verb!PhonoPy! 
software \cite{phonopy}. Large supercells and dense {\bf k}-point grids for Brillouin zone (BZ) sampling were employed. 
Zero-point energy (ZPE) corrections were calculated within the quasi-harmonic approximation \cite{cazorla17,cazorla22}. 
Due to the large number of materials and phases analyzed in this study, thermal expansion effects were disregarded in 
our calculations. 
\\

{\bf First-principles molecular dynamics simulations.}~\emph{Ab initio} molecular dynamics (AIMD) simulations based on 
DFT were performed in the canonical $(N,V,T)$ ensemble (i.e., constant number of particles, volume and temperature). The 
selected volumes were those determined at zero temperature hence thermal expansion effects were neglected. The temperature 
in the AIMD simulations was kept fluctuating around a set-point value by using Nose-Hoover thermostats. Large simulation 
boxes containing $N \sim 200$ atoms were employed and periodic boundary conditions were applied along the three supercell 
lattice vectors. Newton's equations of motion were integrated using the customary Verlet's algorithm with a time step of 
$1.5 \cdot 10^{-3}$~ps. $\Gamma$-point sampling for reciprocal-space integration was employed in most of the AIMD simulations, 
which comprised total simulation times of approximately $20$~ps.
\\

{\bf Crystal structure searches.}~We used the \verb!MAGUS! software (Machine learning And Graph theory assisted Universal 
structure Searcher) \cite{magus} to find new candidate stable and metastable phases for all the He--RG compounds considered
in this study. This crystal structure prediction software employs an evolutionary algorithm augmented with machine learning 
and graph theory to reduce the cost of the geometry optimizations. The crystal phase searches were conducted for structures 
containing a maximum of $6$ formula units.   
\\

\section*{Acknowledgements}
C.C. acknowledges support by MICIN/AEI/10.13039/501100011033 and ERDF/EU under the grants TED2021-130265B-C22,
TED2021-130265B-C21, PID2023-146623NB-I00, PID2023-147469NB-C21 and RYC2018-024947-I and by the Generalitat de
Catalunya under the grants 2021SGR-00343, 2021SGR-01519 and 2021SGR-01411. Computational support was provided
by the Red Española de Supercomputación under the grants FI-2024-1-0005, FI-2024-2-0003, FI-2024-3-0004,
FI-2024-1-0025, FI-2024-2-0006, and FI-2025-1-0015. This work is part of the Maria de Maeztu Units of Excellence
Programme CEX2023-001300-M funded by MCIN/AEI (10.13039/501100011033).
\\

\end{document}